# The IMF impact on the ULF oscillations in the magnetospheric polar cusps

A.V. Guglielmi, O.V. Kozyreva

*Institute of Physics of the Earth RAS, Moscow, Russia*

**Abstract.** Ultra-low-frequency (ULF) oscillations of the magnetospheric cusps are observed permanently in the near-mid sector of auroral oval in the form of so-called IPCL. In this paper, we posed the question: do electromagnetic waves incident on cusp from the foreshock (it is a special region of cosmic plasma existing before the front of the magnetosphere) affect the IPCL regime? We have proposed a method of experimental investigation, which is based on the idea of the foreshock position dependence on the interplanetary magnetic field (IMF) orientation. A hypothesis has been put forward on the existence of a specific effect of the IPCL north-south asymmetry. We tested our hypothesis using IPCL observation data from the Hornsund (Spitsbergen) and Davis (Antarctica) observatories. We found confirmation of the hypothesis at a high confidence level of statistical significance.

## 1. Introduction

The magnetospheric polar cusps are two special areas in the northern and southern hemispheres of the magnetosphere (e.g., [Yamauchi et al., 1996]). They extend from the ionosphere to the magnetopause near the midday meridian. In the projection to the plane of the midday meridian, the cusps resemble horns resting on the Earth at high latitudes with their sharp ends. Observations testify the excitation in cusps of powerful geomagnetic oscillations in the ULF range.

In the literature, it has become a tradition to use the name *Irregular pulsations continuous long*, or abbreviated IPCL for the designation of ULF oscillations of the cusps [Troitskaya, 1985; Troitskaya, Bolshakova, 1977]. This is quite understandable to the researcher of the magnetosphere, and we will use it in this article. But in the interest of geophysicists of other specialties, it is useful to point here to the place of IPCL within the framework of the canonical taxonomy of ULF oscillations.

Let us first recall that the foundations of the systematics of ULF oscillations were laid in the 1960s (e.g., see the review [Troitskaya, Guglielmi, 1967] and the monographs [Jacobs, 1970, Nishida, 1978; Guglielmi, Pokhotelov, 1996]). Pulsations of all types are divided into two classes: Pc (*pulsations continuous*), and Pi (*pulsations irregular*). The type is symbolized as PcN (N=1–5) or PiN (N=1–2), where the number N is the subrange number of the general range of ultra-low frequencies. The Pc class was originally allocated a range of periods from 0.2 to 600 s, and the class Pi, from 1 to 150 s. Soon, however, it became clear that the Pi range should be expanded, and, accordingly, the Pi3 type should be added to denote irregular oscillations in the range 150–600 s (sometimes the upper limit of the range is indicated as 1000 s, or even 1200 s). IPCL refers to the Pi3. It is useful, however, to make one refinement.

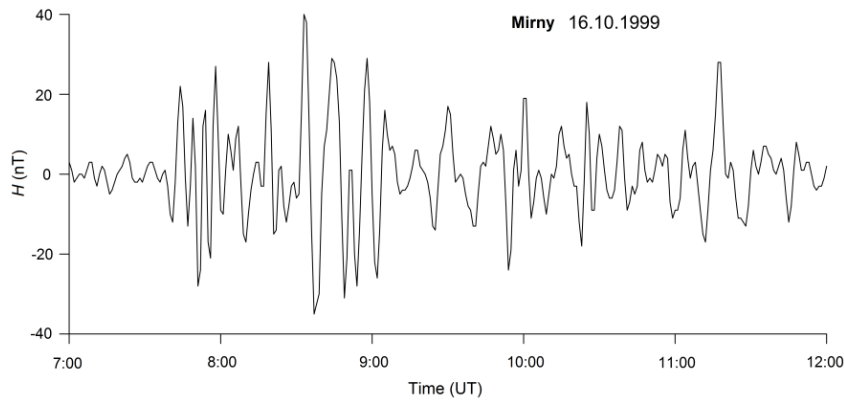

**Fig 1.** The IPCL at Mirny (Antarctica) on October 16, 1999 [Guglielmi et al., 2017].

The abbreviation IPCL indicates that pulsations of this subcategory, while being irregular, are continuous. In these cases, after a number denoting a subrange number, the letter C (from the word *continuous*) is put for greater clarity. Thus, in this article we will talk about the geomagnetic pulsations of Pi3C type. This is an accurate indication of the IPCL within the framework of the systematics of the ULF oscillations. Figure 1 presents an oscillogram of IPCL recorded at Mirny Observatory (corrected geomagnetic coordinates $\Phi= -76.93°$, $\Lambda=122.92°$) [Guglielmi et al., 2017]. The oscillation spectrum is wide. The amplitude of oscillations is variable with typical values from 5 to 10 nT and sometimes it reaches many tens of nT. Seasonal activity of IPCL is characterized by a summer maximum. On average, the amplitude increases with solar wind velocity.

In this work, we investigate the effect of the interplanetary magnetic field (IMF) on the formation of the IPCL north-south asymmetry. It is useful to recall that the question of asymmetry of the ULF geomagnetic oscillations has a long history. The "day-night" asymmetry was discovered long ago: Pc3 (Pi2) pulsations are observed mainly during the day (night) [Troitskaya, Guglielmi, 1967]. The asymmetry "morning-evening" is also well and long-time known: Pi1C are observed in the morning [Kalisher, 1975], and IPDP are observed in the evening [Troitskaya, 1961]. Both these asymmetries are completely due to the asymmetry of the geomagnetic field with respect to the plane of the morning-evening meridian. Along with this, there are two other asymmetries. They are due to the orientation of the IMF. The preferential orientation of the IMF along the Parker spirals leads to the fact that Pc3 are more often observed before the noon than in the afternoon. And, finally, this or that orientation of the IMF with respect to the plane of geomagnetic equator leads to the north-south asymmetry of ULF oscillations.

The north-south asymmetry was observed in the activity of IAR (Pc1–2) at mid-latitude observatory Mondy [Guglielmi, Potapov, 2017], and in the number of occurrence of the discrete signals Pi1B in polar caps [Guglielmi1 et al., 2018]. Asymmetry of a kind was also found in relation to IPCL [Guglielmi1 et al., 2017]. Namely, it was found that in the southern hemisphere (Mirny Observatory, Antarctica), the amplitude of the oscillations increases at a definite orientation of the IMF, and a prediction was made that, when observed in the northern hemisphere, the amplitude of the oscillations will be increased at the opposite orientation. In this paper, we will verify this prediction. Moreover, we will make a statistical check of the effect presence from observations in the southern hemisphere, substantially widening the sample size. Finally, we will analyze the synchronous observations of IPCL at two observatories, one of which is located in the northern hemisphere and the other in the southern hemisphere.

## 2. Research method and data selection

As is well known, before the magnetopause there is the front of shock wave, and before it the so-called plasma-wave foreshock [Russell, Hoppe, 1983]. The foreshock differs from interplanetary plasma in the sense that there is an electromagnetic interaction of two currents here, namely, the flux of the solar wind and a more rarefied flow of those few electrons and ions of the solar wind that are reflected from the shock front. Because of the interaction of two opposing streams the MHD waves are excited in a wide frequency range [Russell, Hoppe, 1983].

MHD waves cross the shock front and fall on the magnetopause. This is the general picture. To understand our method, we must now take into account two important features. First, the plasma-wave foreshock is located asymmetrically relative to the plane of the geomagnetic equator. Secondly, the asymmetry of the foreshock is controlled by the variable orientation of the IMF lines of force.

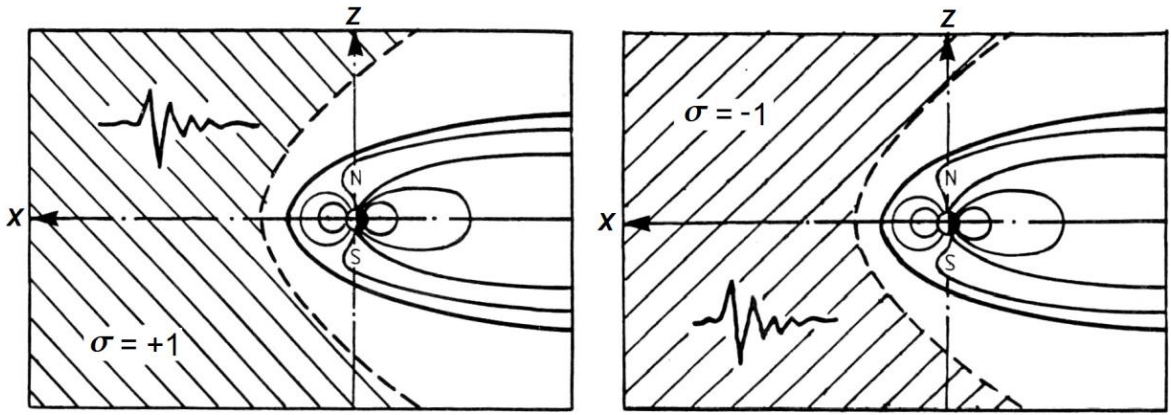

**Fig 2.** Cross-sections of the magnetosphere by the plane of the noon meridian at two IMF orientations [Guglielmi, Potapov, 2017].

We introduce the function *sign* ($B_x \cdot B_z$) and denote it for brevity *sgn*. The argument of our signum function is the product of two IMF components $B_x$ and $B_z$ in the geocentric solar-magnetospheric coordinate system (Figure 2). Omitting the subtleties that can be read in [Guglielmi, Potapov, 2017], we will outline here the most important. If *sgn* = −1, then electromagnetic waves from the foreshock cross the shock front and fall to the southern hemisphere, acting on the estuary of the southern cusp. If *sgn* = +1, then the impact of the waves from the foreshock is subjected to the estuary of the northern cusp. This suggests the procedure for testing the hypothesis of the north-south asymmetry of excitation of the cusps. Namely, we should use *sgn* as the controlling parameter and determine in the experiment whether the IPCL activity in the northern (southern) hemisphere is actually higher for *sgn* = +1 (*sgn* = −1) than for *sgn* = −1 (*sgn* = +1). In other words, if external sources contribute to the excitation of IPCL, then we will observe the antisymmetric reaction of cusps on the switching the sign of *sgn*.

For verification, we chose the Hornsund observatory (corrected geomagnetic coordinates Φ = 74.34°, Λ = 108.21°) in northern hemisphere. In the southern hemisphere, we chose the Davis observatory (corrected geomagnetic coordinates Φ = –76.59°, Λ = 126.23°). Observatories are almost at the same latitudes and at fairly close longitudes. This allowed us to make not only the statistical hypothesis testing separately in the northern and southern hemispheres, but also produce a synoptic IPCL observations at different foreshock orientations.

Selection, processing, and analysis of events have been carried out as follows. By analyzing magnetograms for 1995–1998, we selected events recorded within ±2 h relative to the local geomagnetic noon. As a result, a database was created, consisting of several thousand events. Each event is associated with hourly averages of the three IMF components listed in the OMNI2 catalog [http://omniweb.gsfc.nasa.gov/ow.html]. In this preliminary report, we confine ourselves to the analysis of those events that occurred against a background of weak geomagnetic activity (Kp < 3). A more complete analysis will be presented in another paper.

### 3. Preliminary analysis result

Let $N^+$ ($S^+$) be the amplitude of oscillations in the northern (southern) hemisphere at *sgn* = +1. In a similar way, we denote the amplitude in the northern (southern) hemisphere at *sgn* = −1 as $N^-$ ($S^-$). And let $R^{\pm} = N^{\pm}/S^{\pm}$. We would like to consider the rough numerical characteristics of distributions of the quantities $R^{\pm}$ (mean values and quantiles). Let's start with the events of 1995. When geomagnetic activity is weak (Kp < 3) and IPCL periods ranging from 3 min to 10 min, we have the sample sizes 479 and 501 events for estimate $R^+$ and $R^-$ respectively. For the ratio of amplitudes, we obtained the mean values of $R^+ = 1.34 \pm 0.03$ and $R^- = 1.21 \pm 0.03$. Formally, already at this stage of the analysis, we find a statistically significant distinction between $R^+$ and $R^-$, since the "three sigma" rule is observed. If we summarize the data for all four years, then $R^+ > R^-$ at the level of five sigma. In other words, the confidence level of statistical significance satisfies even that rigid criterion, which is accepted in the physics of elementary particles. The inequality $R^+ > R^-$ testifies to our hypothesis about the north-south asymmetry of excitation of cusps, which depends on the orientation of the IMF line of force relative to the plane of the geomagnetic equator.

We could be now confidently state that a new property of IPCL is discovered, but there is a nuance: The first, third and fifth sextiles of the distributions are 0.7, 1.2 and 2.0 for sgn = +1, and 0.7, 1.1 and 1.7 for sgn = −1. We clearly see that our distributions are slightly sloped to the left and, therefore, they are not accurately described by the normal distributions. We applied the nonparametric Kolmogorov-Smirnov test to compare two sample distributions and found that small deviations from normal distributions did not affect the conclusion that $R^+ > R^-$. The question of the shape of amplitude distributions, interesting in itself, we plan to study in more detail. Apparently, we are dealing here with the gamma distributions $\Gamma_{\alpha,2}(N^\pm)$ and $\Gamma_{\alpha,2}(S^\pm)$.

**Table.** Mean values of $R^\pm$ in the period range of 10–20 minutes.

| Year | $R^+$ | $R^-$ |
|------|-------|-------|
| 1995 | 1.47  | 1.22  |
| 1996 | 1.43  | 1.23  |
| 1997 | 1.26  | 1.20  |
| 1998 | 1.34  | 1.27  |

A similar result was obtained from data on IPCL oscillations in the range of 10–20 min (see the Table). Apparently, we can recognize as highly plausible the hypothesis about the effect of IMF orientation on the vibrational activity of magnetospheric cusps.

### 4. Discussion

We calculated the $R^\pm$ values based on measurements of the IPCL amplitudes at the Hornzund and Davis observatories. It would seem that by averaging over all orientations of the IMF, we must obtain the unit (with accurate to a random error in the measurement of the amplitudes). Moreover, in an ideal experiment to test our hypothesis about the north-south asymmetry of IPCL, we would have to obtain not only two inequalities $R^+ > R^-$ and $R^+ > 1$, but also an inequality $R^- < 1$ under the additional condition $R^+ + R^- = 2$ (the latter up to a measurement error). We have seen that the inequalities $R^+ > R^-$ and $R^+ > 1$ are indeed satisfied. But the inequality $R^- < 1$ and the condition $R^+ + R^- = 2$ are not satisfied in a real experiment. The reason for this is simple, and it lies in the fact that amplitude measured at the observatory Hornsund, systematically exceeds the amplitude measured at the observatory Davis.

Systematic errors occur in every measurement. Experience shows that, unlike random errors, it is not easy to take systematic errors into account, and it is almost impossible to eliminate them in many cases. Where in our particular case there was an unaccounted error, remains incomprehensible. The source of the error may be the difference in the characteristics of the magnetometers located at the Davis and Hornsund. Now, more than 20 years after IPCL registration, this can only be guessed. A partial role may also be played by the incomplete magnetic conjugation of observatories. A permanent source of a small error in the magnetometric investigations of the alternating electromagnetic field is also a distinction in the surface impedances of the Earth's crust at the Davis and Hornsund. However, under any circumstances, a systematic error leading to a violation of the inequality $R^- < 1$ can not in any way affect the radical conclusion that $R^+ > R^-$ takes place at a high level of statistical significance.

What is the meaning of the inequality $R^+ > R^-$? We believe that the idea of the permanent effect of the plasma-wave foreshock on the vibrational activity of magnetospheric cusps lies at the basis of a plausible interpretation of the inequality $R^+ > R^-$. The foreshock affects either the northern or the southern cusp, depending on the orientation of the IMF.

## 5. Conclusion

The preliminary result of our work is the following:

1. The database for statistical and synoptic studies of the effect of the interplanetary magnetic field on the vibrational activity of magnetospheric cusps has been created.

2. The methodology is proposed for investigation the dependence of the magnetospheric cusps oscillations on the orientation of the foreshock.

3. The hypothesis of existence of the specific north-south asymmetry of magnetospheric cusps oscillations has been confirmed.

A more complete analysis of the north-south asymmetry of magnetospheric cusps oscillations will be presented in a separate paper.

*Acknowledgements*. We express our deep gratitude to B.I. Klain and A.S. Potapov for numerous discussions of the problem. This work was partially supported by the Program 28 of the Presidium of RAS and RFBR project # 16-05-00056.